\documentclass[preprint]{vldb}
\usepackage[utf8x]{inputenc}
\usepackage{subcaption}
\usepackage{graphicx}
\usepackage{pslatex}
\usepackage{amssymb}
\usepackage{verbatim}
\usepackage{fancybox}
\usepackage{amsmath}
\usepackage{amsfonts}
\usepackage{amsmath}
\usepackage{algorithm}
\usepackage{algorithmic}
\usepackage{subfig}
\usepackage{graphicx}
\usepackage{comment}
\usepackage{multirow}
\usepackage{color}
\usepackage{subcaption}
\usepackage{flushend}
\usepackage{lmodern}

\toappear{} 

\newdef{definition}{Definition}

\definecolor{orange}{RGB}{255,127,0}

\newcommand{\remark}[5]{{ \colorbox{#1}{\color{#2} $\triangleleft$ \textbf{#3}} } {\color{#1}\ifthenelse{\equal{#5}{}}{}{#5:} } #4 { \colorbox{#1}{\color{#2} $\triangleright$}}} 

\title{Diversification Based Static Index Pruning\\
\LARGE Application to Temporal Collections}

\numberofauthors{3}

\author{
\alignauthor
Zeynep Pehlivan\\
\affaddr{LIP6, University P. and M. Curie}\\
\affaddr{4 place Jussieu 75005, Paris, France}\\
\email{\small zeynep.pehlivan@lip6.fr}
\alignauthor
Benjamin Piwowaski\\
\affaddr{LIP6, University P. and M. Curie}\\
\affaddr{4 place Jussieu 75005, Paris, France}\\
\email{\small benjamin@bpiwowar.net}
\alignauthor 
St\'{e}phane Gan\c{c}arski\\
\affaddr{LIP6, University P. and M. Curie}\\
\affaddr{4 place Jussieu 75005, Paris, France}\\
\email{\small stephane.gancarski@lip6.fr}
}

\begin{document}
\maketitle

\begin{abstract}
Nowadays, web archives preserve the history of large portions of the web. 
As medias are shifting from printed to digital editions, 
accessing these huge information sources is drawing increasingly more attention from national and international 
institutions, as well as from the research community. 
These collections are intrinsically big, leading to index files that
do not fit into the memory and an increase query response time. Decreasing the index size is a direct way to decrease this
query response time.

Static index pruning methods reduce the size of indexes by removing a part of the postings. 
In the context of web archives, it is necessary to remove postings while preserving the \textit{temporal 
diversity} of the archive. 
None of the existing pruning approaches take (temporal) diversification into account. 

In this paper, we propose a \textit{diversification-based static index pruning} method. It differs from the existing pruning approaches by integrating diversification within the pruning context. 
We aim at pruning the index while preserving retrieval effectiveness and diversity
by pruning while maximizing a given IR evaluation metric like DCG.
We show how to apply this approach in the context of web archives.
Finally, we show on two collections that search effectiveness in temporal collections 
after pruning can be improved using our approach rather than diversity oblivious approaches.

\end{abstract}


\section{Introduction}

In the past ten years, content producers began shifting to digital media. Consequently, libraries, that aim at preserving 
cultural artifacts and providing access to them, have begun to shift towards this new media (e.g. archive.org). On the web, preservation
implies crawling regularly changing web pages, generating temporal collections called \textit{web archives}.

The major challenges with these temporal collections are the efficient storage and retrieval of information. Querying these
collections requires the use of temporal queries that combine standard keyword
queries with temporal constraints. The results of such queries can be obtained by using time-aware ranking models that rank the documents based on temporal and textual similarities 
\cite{gurrin_language_2010, nattiya2010}
or by filtering out the topically relevant documents with time intervals \cite{nutchwax}.

To efficiently process these queries, information systems rely on an inverted index.
It provides a map between terms and their occurrences in the documents.
It consists of a collection of postings lists, each associated with a unique term in the collection.
Each posting holds some information, typically a document identifier and the frequency of the term in that document.

When querying web archives, the same data structures are used.
NutchWAX, an extension to Nutch that uses inverted index data structure,
supports full-text search for the Finnish Web Archive for 148 millions of contents (e.g. html pages, pdfs, images etc.), Canada WA for 170 millions of contents,
Digital Heritage of Catalonia for 200 millions of contents \cite{gomes_survey_2011}. As they store the evolving Web, temporal collections require big disk spaces, which in turn leads to big
index files that
do not fit into the memory. Consequently,
the search engine needs lots of disk-accesses, increasing query response time.

To solve this problem, pruning techniques, that compute the results without scanning the full inverted index, were proposed. \textit{Index pruning} removes postings
that are less likely to alter the ranking of the top-k documents retrieved by query execution.
Index pruning can be dynamic or static. Dynamic pruning prunes during query processing by deciding which postings should be involved in the ranking process, and whether the
ranking process should stop or continue. 

On the other hand, static pruning reduces the index size by removing postings from the index offline, independently from any query. It is a lossy index compression technique because it is not
possible to bring the compressed data back to its
exact uncompressed form. The challenge is to decide which information to prune without decreasing too much the retrieval effectiveness.
Obviously, for temporal collections, this decision should take the temporal dimension into account.

However, none of the existing pruning techniques deals with the problem of temporal collections and using existing pruning techniques may lead to loose postings from different time periods.
Consider the example of the temporal query ``Iraq War in  1991''. If the documents mentioning ``Iraq War in 2003'' have higher scores than the documents mentioning ``Iraq War in 1991'',
the index after pruning will
have less or no relevant documents for the first temporal query. Ideally, we would like the pruning method to keep the relevant
documents for different time periods. Pruned indexes should preserve the temporal diversity of postings.

This issue is related to the problem of \textit{search result diversification} that aims to improve user satisfaction according to different information needs. Search results should be optimized to contain
both relevant and diverse results.
A number of search result diversification methods are proposed \cite{welch_search_2011,agrawal_diversifying_2009,carbonell_use_1998}.
We base our work on coverage based diversification, which is an optimization problem of an objective function related to both relevance and diversity of the results. In our case, the objective function is an IR metric adapted to diversification. 

As diversification is computationally intensive, we chose to follow a static pruning approach where postings are pruned off-line. 
To the best of our knowledge, static index pruning and search result diversification were never studied together.
In this paper, we propose a new approach that we name
\textit{Diversification-based static index pruning}.

Although, in our context we focus on temporal coverage, the method can also be applied to non-temporal collections
with different diversification categories. The contributions of this paper are the following:

\begin{itemize}
\item We take into account diversification in the context of index pruning and propose a variation of the standard greedy algorithm.
\item We show how to apply this method to temporal diversification index pruning by defining the temporal aspects associated to a query.
\item We perform experiments on two datasets that were modified for the task of retrieval in web archives and show that our approach is stable and works better than other index pruning stategies.
\end{itemize}


The rest of this paper is organized as follows: Section \ref{sec:relatedwork} puts our work into context with related research. In Section \ref{sec:divbased}, we introduce our model
for Diversification based static index pruning. Section \ref{sec:tsip} shows how to apply our approach to temporal collections. 
Section \ref{sec:experiments} describes the series of experiments and analysis. We conclude and discuss future work in Section \ref{sec:concl}.


\section{Related Works}
\label{sec:relatedwork}

In ad-hoc information retrieval (IR), the goal is to retrieve
most relevant documents according to a topic or \emph{query}. In 
order to determine a ranking of documents, a score of relevance is computed for each document.
Indexes are used to compute these scores efficiently at query time. 
As we aim to keep both most relevant and most diverse postings in the index, 
our work is related to two IR domains, namely Static Index Pruning and Search Results Diversification. 
Both domains were independently studied, and we discuss them separately in the next two sections. 

\subsection{Static Index Pruning}

Static index pruning reduces the index size by discarding the postings that are less likely to affect the retrieval performance. 
%
The seminal work of Carmel et al. \cite{carmel_static_2001} was based on two simple ideas: each posting is associated
to a score, and a threshold on this score is defined to filter out a part of the postings. 
The main differences between all the approaches that came latter comes from the way the score and threshold are defined.

In the case of Carmel  \cite{carmel_static_2001}, the score is the TF-IDF value associated with a posting.
De Moura et al. \cite{moura_locality-based_2008} proposed to use term co-occurrence
information.
More recently, Blanco et al. \cite{Blanco_2010} used the 
the ratio of the probability that a document is relevant to a query over 
the probability that a document is not relevant to a query (odd-ratio) 
is used as a decision criteria. We follow 
a similar approach where every term in the vocabulary is considered as a single term query.
Chen et al. \cite{CIKM_2012} 
proposes to use a measure related to the contribution of a posting to the entropy of 
the distribution of documents given a term.   
Thota et al. \cite{Ecir_2011} compare the frequency of occurrences of a word in a given 
document to its frequency in the collection and use the two sample two proportion statistical test to decide whether to keep a posting or not.

With respect to the thresholds, Carmel et al. \cite{carmel_static_2001} defines a threshold that depends on each term in a way that preserves at least the top-$k$ postings of this term (when ranked by their score).
A limitation of this approach is that it may retain too many postings for unimportant terms (those that will not be important when ranking documents). 
In this case, it is possible to use a global threshold (\textit{uniform pruning}) for all the terms like in \cite{Ecir_2011, CIKM_2012}.

There are other approaches to pruning that are not variations of Carmel's one \cite{carmel_static_2001}.

Instead of ranking the postings per term, postings can be ranked by document.
B\"uttcher and Clarke \cite{buttcher_document-centric_2006} introduced another pruning method that removes the posting lists based on the documents. For each document D in the corpus, 
the terms of the document are ranked
based on their contribution to the Kullback-Leibler divergence between the
distribution of terms within the document and the collection. Then, only the postings
for the top-k terms in the document are kept in the index. 

Another way to prune is to remove whole postings list for a subset of terms or documents.
Blanco et al. \cite{blanco_static_2007} proposed to remove entire posting list based on informativeness value of a term, in particular inverse document frequency (idf) 
and residual inverse document frequency (ridf), and two methods based on Salton's term discriminative model \cite{salton1975theory}, 
which measures the space density before and after removing a dimension (term).
This approach may remove entire terms from the index and should be reserved to a small subset of the vocabulary. 
Finally, it is possible to fully remove a document from the postings -- this is related to the problem of how retrievable a document is \cite{Azzopardi2008Retrievability}.
Zheng and Cox \cite{zheng_entropy-based_2009} suggested an entropy based document pruning
strategy where a score is given to each each document based on the entropy of each of its terms. Documents
below a given threshold are then removed from the index. 
These types of approaches are orthogonal to ours and could thus be implemented on top of it.

All the works mentioned above focus on different pruning techniques.
In our work, we depart from the above approaches since (1) we use an IR metric
to select the set of postings that we keep in the index and (2) we try to preserve
diverse postings -- where diversity is dependent on the task at hand, here searching
web archives.

\subsection{Search Results Diversification}
Result diversification aims to return a list of documents
which are not only relevant to a query but also cover many subtopics of the
query. The approaches can be classified as \textit{explicit} and \textit{implicit} \cite{Ahlers2012LocalWebSearch}. 

Implicit methods consist in evaluating a similarity measure between documents and to use this information to select diverse documents by discarding too similar documents.
One of the 
earliest approaches \cite{carbonell_use_1998}, \textit{Maximal Marginal Relevance} (MMR),
select documents by computing a score which is a linear combination of a novelty score (dissimilarity) and the original relevance score.
The various approaches based on MMR differ mostly by
how the similarity between documents is computed \cite{mmrzhai,mmrjun}.

Explicit methods suppose that the different aspects of a query can be computed along with the extent to which a document is relevant to each of these aspects.
The most representative work based on explicit methods is that of Agrawal et al. \cite{agrawal_diversifying_2009} where the authors propose to maximize the probability of 
an average user finding at least one relevant result. They assume that an explicit list of subtopics is available for each query.
Recently, Welch et al. \cite{welch_search_2011}
observed that this approach falls into random document selection after choosing one document for each subtopic.

In the context of index pruning, it is important to reorder 
all the documents based on diversification.
A way to avoid this random document selection is to use a criterion based on
the maximization of an IR metric. This type of criterion is used to learn to rank in ad-hoc IR \cite{dcgapproximation,Climf} and has recently been applied to 
diversification \cite{dcgapproximation}. In this latter work, the authors 
use a performance measure based on discounted cumulative gain, which defines the usefulness
(gain) of a document depending on its position.
Based on this measure, they suggest a general approach to
develop approximation algorithms for ranking search results
that captures different aspects of users’ intents. However, no experiments were conducted.

Coverage based diversification corresponds well to our problem of finding temporally diverse results since we can define temporal aspects explicitly as shown in Section \ref{sec:timewindows}.
We want to keep in the index the results that cover many 
different aspects (e.g. temporal aspects) for the given query. Our work
is related to \cite{dcgapproximation} in that we optimize a IR metric. However,
we use it for index pruning and define algorithms to perform this optimization efficiently
along with a definition of temporal aspects related to a given user query.




\section{Diversification Based Static Index Pruning}
\label{sec:divbased}
In this section, we propose a method for static index pruning
that takes the diversification of results into account.
Existing static index pruning methods can discard relevant documents for some aspects of a query (e.g. subtopics)
as they do not take the result diversification into account.
In this section, we propose a method for static index pruning by taking into account the diversification of results. 

\subsection{Problem Formulation}
\label{sec:problemdef}
In order to define a criterion for diversification-based static index pruning, we need to know which queries might be asked by users, which aspects are associated to the queries and how
documents are linked to these aspects (i.e. how relevant is the document with respect to this query aspect).
We define Q as a set of queries that a user can ask, $W_q$ as the set of aspects of a query $q\in Q$. We show in Section \ref{sec:querygeneration}, 
how we approximate the distribution over queries $Q$ and in Section
\ref{sec:tsip}  the distribution over aspects $W_q$.

In order to define a criterion, we chose to follow the approach taken by some works on learning to rank (learning the parameters of a search engine) 
that try to optimize directly an IR metric like in \cite{irlearning1,irlearning2}.
This approach is interesting, since it directly tries to optimize a measure which 
is related to the user satisfaction, and on the other hand, it allows to easily integrate our diversification requirements.

We can now define the criterion that we want to optimize.
Given a set of postings $D$, an evaluation metric M, a number of postings to preserve  $k$, we would like to find a set of
of postings  $S^*$ that maximizes M:

\begin{equation}
 S^* = {\arg\max}_{S\subseteq D,|S| = k} \mathbb{E}(M | S)
\label{eq:argmax}
\end{equation}

If we suppose we can estimate the probability that each individual query $q\in Q$ is asked,
we can compute this expectation as a sum over all the possible queries. This gives:
\begin{equation}
\label{eq:expM}
\mathbb{E}(M|S) = \sum_{q \in Q} P(q)  \sum_{w \in W_q} P(w|q) \mathbb{E}(M|q,w,S) 
\end{equation}
where P(q) is the probability of a user issuing query $q$  and $P(w|q)$ is the distribution on the aspects of query $q$ with $ \sum_{w \in W_q} P(w|q) = 1$. In practice, we assume that the queries are reduced to one term
(Section \ref{sec:querygeneration}) and define what (temporal) aspects a query $q$ is associated with
by different means (Section~\ref{sec:timewindows}).


\subsection{Optimization}
\label{sec:optimization}

We now turn to the problem of 
optimization of Equation \ref{eq:argmax}, which is known to be NP-hard \cite{agrawal_diversifying_2009,carterette_analysis_2011}. 
However, there are known algorithms that give in practice good approximations. 
In our work,
we use the standard \emph{greedy algorithm heuristic} that we describe in this section.

This heuristic is based on the submodular nature of the function given in Equation \ref{eq:argmax}.
\begin{definition}
\label{def:submodularity}
Consider a set X and a function \textit{f} defined on $2^X$. \textit{f} is said to be \textit{submodular} if for any two subsets A and B in $2^X$, the following holds:

\[  f(A)+ f(B) \geq f(A \cap B) + f(A \cup B) \]
\end{definition}

Intuitively, a submodular function satisfies the economic
principle of diminishing marginal returns, i.e., the marginal
benefit of adding a document to a larger collection is less
than that of adding it to a smaller collection \cite{agrawal_diversifying_2009}. Equation \ref{eq:expM} has been shown to be submodular for metrics like DCG \cite{dcgapproximation} that we use in our work.
 
The greedy algorithm is the most commonly used 
since it offers guarantees on the objective function value and works well for a wide range of problems. It chooses the best solution that maximizes (or minimizes) the objective function (also known as marginal 
function or cost function) at each step, independently from the future choices. Finally, all the best local solutions
are combined to get the global optimal/suboptimal solution. As proved in \cite{nemhauser}, the greedy algorithm ensures that the value of the objective function for the approximation is at least (1-1/e) times the optimal one.

We adapt the greedy algorithm to Equation \ref{eq:argmax}, as shown in Algorithm \ref{algo:greedy}. 
The greedy algorithm uses two sets: X for available items and S for selected items. S can be initialized with some items or can be empty.
Then, the algorithm iterates over X and adds the best element (according to the objective) to S until $|S| = k$.
$p_i$ that has the highest value for objective function is chosen. 
\[  p_i = {\arg\max}_p\mathbb{E}(M | S_{i-1} \cup \{p\})\]

Intuitively, the documents that can provide coverage 
for many query aspects should be placed in the beginning of the ordering so as to maximize the objective function.

\begin{algorithm}
\caption{ \textit{Diversify} - Greedy Algorithm}
\label{algo:greedy}
\begin{algorithmic}[1]
\REQUIRE $k$ (target), $S$ the set of all postings, $f(S^\prime)=\mathbb{E}(M|S^\prime)$  
\ENSURE The set of optimal postings $S^*$
\STATE  $S^* = \emptyset $
  \WHILE{ $|S^*| < k$ }
      \STATE $p^* \leftarrow \mathop{\arg\max}_{p\not\in S} f(S\cup \{p\})$
      \STATE $S^* \leftarrow S^* \cup \{p^*\}  $
\ENDWHILE
\RETURN $S^*$
\end{algorithmic}
\end{algorithm}

\subsection{Metric family}
\label{sec:metric}

We use the DCG measure (Discounted Cumulative Gain) \cite{manning2008} as our metric M 
since it is a well-established IR metric that allows an efficient pruning algorithm (Section \ref{sec:pruning})
to be defined. DCG can be written as the sum of gain and discount functions as follows: 
\begin{equation}
 DCG = \sum_{j\ge1} c(j) g(d_j)
\label{eq:dcg}
\end{equation}
where $c(j)$ is a decreasing discount factor\footnote{Usually set to $1 / \log( 1 + j)$ as in this paper}, and $g(d_j)$ is the gain function for a document $d_j$, representing the value of the document for the user. This value is usually defined as a function of the relevance of the document, where the relevance can take several value ranging from 0 (not relevant) to 5 (excellent result).
In the following, we suppose that $M$ belongs to the DCG family.

By choosing 
a decaying function as a discount function, DCG focuses on the quality of the top-k results. Intuitively, in the context of pruning, choosing a metric M that has a bias towards top ranked results is interesting since this will help to establish a balance between relevance (adding one more result dealing with the same query aspect) and diversity (adding one more result dealing with another query aspect).

The definition of DCG as a sum over ranks allows to get a closed
form formula where the relevance of each document is explicit.
Starting from Equations \ref{eq:dcg}, we get: 
\begin{equation}
 \mathbb{E}[M|q,w,S] = \sum_{j\ge1} c(j)  \mathbb{E}[g(d_j)|q,w,S]
\label{eq:expecteddcg}
\end{equation}

Since we don't know the relevance of the documents to the query $q$,
we define the expectation $\mathbb{E}[g(d_j)|q,w,S]$ of the gain as the probability of relevance 
$P(d_j|w,q)$ of document $d_j$ for query $q$. The probability is directly given by an IR model and it is equal to zero if 
w is not an aspect of $d_j$.  Note that we could relax this latter requirement
by allowing documents to be more or less relevant given the aspects, but
we did not consider this in this work.
From Equations \ref{eq:expM} and \ref{eq:expecteddcg}, we get:
\begin{equation}
 \mathbb{E}[M|S] = \sum_q P(q) \sum_{w\in W_q} P(w|q) \sum_{j\ge1} c(j)  P(d_j|q,w,S)
\label{eq:expecteddcg2}
\end{equation}

\subsection{Query Generation Model}
\label{sec:querygeneration}

We now turn to the estimation of the distribution over the queries $q\in Q$.
As we work on static index pruning which is applied off-line, we do not have any knowledge over Q. 
As an approximation like in \cite{Blanco_2010}, we assume that each query consists of one single term. More sophisticated strategies
would involve using query logs and/or using co-occurrence information, but we leave
this for future work.

Starting from Equation \ref{eq:expecteddcg2}, we get:
\begin{equation}
 \label{eq:dcg_per_term}
 \mathbb{E}(M|S) = \sum_t P(t) \sum_{w \in W_t} P(w|t) \sum_{j\ge1} c(j) P(d_j|t,w,S)
\end{equation}

As discussed in Section \ref{sec:optimization}, it is shown that the greedy algorithm still
 achieves (1-1/e) approximation even in the more general settings of using a submodular 
function with a logarithmic discount function \cite{dcgapproximation}.

We hence use a greedy algorithm to find the best posting to add to the current postings
list. The above equation allows us to define an efficient algorithm to find this posting:

$$P(d_j|t,w,S)=
\begin{cases}
    P(d_j|t) & \mbox{if } p_{t,d_j} \in S \mbox{ and } w \mbox{ matches } d_j\\
    0 & \mathrm{otherwise}
\end{cases}$$

\newcommand{\dpr}{{d^\prime}}

We want to choose the posting that will maximize the criterion of
 Equation \ref{eq:expecteddcg2}. We first define
$\Delta(S,p_{t,d})$ as the change in the value of the criterion that
we would get by adding to S the posting $p_{t,d} \not\in S$ for a document $d$ and term $t$.
Let $l=(d_{t,w,j})$ be the list of $N_{t,w}$ documents whose postings $p_{t,d_{t,w,j}}$ belong to
the already selected postings, ordered by decreasing probability of
relevance\footnote{we discard those with a probability 0 as they will not change the criterion} 
$P(d_{t,w,j}|t,w)$, and let $r_{t,w}$ be the first rank where $p(d|t,w)>p(d_{t,w,r}|t,w)$.

In this case, adding $p_{t,d}$ to the posting list will change the rank of all the documents
after rank $r_{t,w}$. Furthermore, the gain of the document $d$ will be added to the DCG value.
All the above allows us to define the increase in DCG as:
\begin{equation}
\label{eq:posting_delta}
\begin{split}
\Delta(S,p_{t,d}) = 
   P(t) \sum_{w \in W_t} P(w|t) 
   \times \Bigl[ c(r_{t,w}) P(d|t,w) \\
      + \sum_{j=r_{t,w}}^{N_{t,w}} (c(j+1)-c(j)) P(d_{t,w,j}|t,w) \Bigr]
\end{split}
\end{equation}

A naive approach to selecting the best posting -- e.g. by computing explicitly
the list of documents $l$ would be too computationally expensive. We propose the algorithm \ref{algo:greedyterm}
that computes the next best posting to select; the overall
complexity is given by $O\left( \sum_{t\in V} |P_t|^2 |W_t| \right)$,
i.e. the complexity is quadratic with respect to the size of the posting list for a term $t$ and linear
with respect to the number of considered aspects $W_t$.

The algorithm \ref{algo:greedyterm} shows how to get the next best posting for a given term.
At the first round, we need to compute this for each term, but then, for each
round of the greedy algorithm, we just need to compute the next best posting for
the term corresponding to the posting we just selected.
The proposed algorithm is bottom-up, that is, for each term we start from the 
less relevant documents first, allowing us to optimize the computation of the
last sum in Equation \ref{eq:posting_delta}.

We take as inputs a list of tuples, $P_t$, a set of selected postings $S$, a set of aspects $W$ and a mapping $m$ from the documents to the powerset of aspects $W$. 
$P_t$ consist of tuples $( d_i, p(t|d_i))$ that contain document ($d_i$), its score based on the probability of relevance $p(t|d_i)$. $S$ is used to track which postings are already selected.
The set of aspects $W$ is a set of tuples $(t_{w}, p_{w})$  where 
$t_{w}$ is the number of selected postings and $p_{w}=p(w|t)$ gives the distribution over the aspects for the given term. Aspects have also associated properties (like the times windows in web archives)
but this is not relevant for the algorithm.
In the following, we access parts of the tuples by using them as functions, e.g $d_i(P_t)$ refers
to the $d_i$ entry of the $i^{th}$ element of the list.

In the algorithm, each aspect $w$ in $W$ is associated to a state given
by $r_w$ and $\Delta_w$, where
$r_w$ holds rank position of the current $p_{d_i,t}$ in the list
of results associated to aspect $w$. $\Delta_w$ holds the value of the last
sum in Equation \ref{eq:posting_delta} up to rank $r_w$.

The core of the algorithm is the computation of the Equation \ref{eq:posting_delta} (lines 10-16) which combines 
the relevance score of the document $d_i$ with respect to $p(t|d_i)$ and a diversity score.  Starting from the bottom of the list $P_t$, 
for each document, we traverse each aspect associated to this document. 
If the document is not selected in the previous execution of the algorithm, 
its score is calculated according 
to Equation \ref{eq:posting_delta} (line 15) by adding the current $\Delta_w$ to the gain for the document. Otherwise, $\Delta_w$ is updated (line 12) and the current rank $r_w$ is decremented.
When we have checked all the list, we can return the best posting along with
the change in the criterion (up to the scaling by $p(t)$).


\begin{algorithm}[t!]
\caption{NextBest($P_t, S, W, m$): Get the next best posting for term $t$}
\label{algo:greedyterm}
\begin{algorithmic}[1]
\REQUIRE $P_t$ list for term $t$ $(d_i, p(t|d_i), t_i)$ ordered by decreasing $p(t|d_i)$\\
 $S$ the list of selected postings \\
 $W$ sets of aspects as tuples $(t_w, p_w)$\\
 $m$ a mapping between documents and a subset of $W$
\ENSURE Next best posting for term $t$ along with the delta in the criterion
  \STATE maxdelta $\leftarrow$ 0
  \STATE maxdoc $\leftarrow$ 0
  \STATE $r\leftarrow$ empty map, $\Delta \leftarrow$ empty map
  \FOR{$w \in W$}
	  \STATE $ r_{w}  \leftarrow t_{w}$
	  \STATE $\Delta_w \leftarrow 0$
  \ENDFOR
  \FOR{$i = |P_t| \to 1$}
      \STATE score = 0
      \FORALL { $w \in m(d_i)$ }
	  \IF{$p_{t, d_i} \in S$}
		\STATE $\Delta_w \leftarrow \Delta_w  + (c(r_{w+1}) -  c(r_w))\times p(t|d_i) $
		\STATE $r_w  \leftarrow r_w - 1$
	  \ELSE
		\STATE $ score   \leftarrow score + ( \Delta_w  + c(r_w+1) \times p(t|d_i))\times p_w$
	  \ENDIF
      \ENDFOR
      \IF{ $ score > maxdelta$ }
	\STATE maxdelta $\leftarrow$ score
	\STATE maxdoc $\leftarrow d_i$
      \ENDIF
  \ENDFOR
\RETURN $(i, \mathrm{maxdelta})$
\end{algorithmic}
\end{algorithm}


\subsection{Pruning}
\label{sec:pruning}
There are different ways to estimate the probability $P(t)$ that should reflect
how likely a user is to issue the query $t$. We could assume that 
$P(t)$ is uniformly distributed, but preliminary experiments have
shown that this did not work well. Another option would be to use query logs to estimate 
the probability of a term to occur in a query. However, query logs (with non anonymized queries)
are not publicly available. 

In this work, we circumvent the problem of estimating $P(t)$
by optimizing each term independently of each other. This amounts at
setting $P(t)$ such that we preserve a pre-determined number of postings
for each term. 

Algorithm \ref{algo:prune} gives the pseudo-code following this approach.
For a given index with vocabulary $T$, we compute the top-k best
postings by calling \textit{NextBest} defined in the previous section.
After each call, we update the aspects where the selected posting appear
to increment the size of the list associated with the aspects.
We keep in the index the top-k postings and discard the other postings.
 If k is the same for all terms in the vocabulary, it is called \textit{uniform top-k pruning}. 

Our approach can be also used with other pruning methods like the ones proposed in \cite{CIKM_2012,Ecir_2011} by defining a global threshold $\epsilon$ rather than choosing top-k for each term. In that case, Algorithm 3 
would need to be updated, but we don't detail this in this paper.


\begin{algorithm}
\caption{ \textit{Diversified Top-K Pruning}}
\label{algo:prune}
\begin{algorithmic}[1]
\REQUIRE for each term, the desired number of postings ${k_t}$, the order list of postings ${P_t}$, the set of aspects $W_t$ and the mappings $m_t$ between documents and the subsets of aspects.
\ENSURE $P$ the optimal set of postings
\STATE 	$P \leftarrow \emptyset$
\FORALL{ $t \in T$}
	\STATE $S_t \leftarrow \emptyset$
	\WHILE {$|S_t| < k_t $ }
		\STATE $(i, \Delta) \leftarrow NextBest(P_t, S_t, W_t, m_t)$
	    \STATE $S_t \leftarrow  S_t \cup \{p_{t, d_i(P_t)}\}$
		\FOR{$w \in m_t(d_i)$}
			\STATE $r(w) \leftarrow r(w) + 1$
		\ENDFOR
	\ENDWHILE
	\STATE $P \leftarrow P \cup S_t$
\ENDFOR
\RETURN  P
\end{algorithmic}
\end{algorithm}


\section{Temporal Static Index Pruning}
\label{sec:tsip}

In this section, we show how to use diversification based static index pruning to ensure temporal diversity in temporal collections.
Documents and queries are associated to temporal aspects (i.e. a series of time spans). We use \cite{gurrin_language_2010} as based time model to 
define temporal aspects and the matching the aspects between documents, queries and the aspects. 

We define a temporal aspect as a \textit{time window}, i.e. a time interval.
A query (or a document) can refer to a set of temporal aspects $W_q$ (or $W_d$).
While we suppose that the set of temporal aspects for documents is known (e.g.
the validity time interval in web archives), this is not the case for queries since we do not know them in advance.

More precisely, given a term t, we now need to determine what are its different temporal aspects $W_t$ and their importance by defining a distribution over $W_t$ as discussed in Section \ref{sec:problemdef}.

In the following, we define three different strategies. The first two are based on fixed-size time windows spanning the entire time range. The only parameter is the duration (size) of time windows.
However, this strategy may fail when the term is related to events whose temporal span vary. For example, the term ``olympics'' depend on the year, on the other hand, the term ``earthquake'' 
can take a place any time and its time span can vary based on the collection and the magnitude of the earthquake. 
To handle this case, a dynamic strategy based on a mixture of gaussians is proposed. 

In this section, we first present more formally the temporal expressions, and how to match
a document to a query, following the work of Gurrin et al. \cite{gurrin_language_2010}. 
We then explain in detail our three different strategies to define the temporal aspects $W_t$ of a term.

\subsection{Time Windows}
\label{subsec:time}

A discrete time model is used to represent the time, according to \cite{gurrin_language_2010} where a temporal expression T is represented as
\[ T = (b,e) \]
where b and e are respectively the start and end date of the time window. Each is as a time range with a lower bound and upper bound, e.g. $b=[b^l,b^u]$.

When the temporal expression is uncertain (e.g ``in 2013''), this time representation allows to capture the inherent uncertainty -- e.g. $T=$([1/12/2013, 31/12/2013], [1/1/2013, 31/12/2013]) -- since we only know the start and end date are in 2013).

While filtering the temporal query results, we need to find if $d_{time}$ intersects with $q_{time}$. The intersection operation correspond to a time window where the start (resp. end) date is the intersection of the start (resp. end) date ranges. More formally, let $l=\max(b_1^l,b_2^l)$ and $u=\min(b_1^u,b_2^u)$. If $l\le u$, then the start date of the intersection is the range $[l,u]$, otherwise it is the empty set -- and likewise for the end date. 

\subsection{Document and Query Model}
\label{sec:docqmodel}

A document $d$ in the temporal collection $D$ consists of two different parts: a textual part, denoted $d_{text}$, and a temporal part, denoted $d_{time}$. 
As usual in IR, $d_{text}$ is represented as a bag of words. Following \cite{gurrin_language_2010}, $d_{time}$ is represented as a bag of temporal expressions (e.g publication date)  as defined in Section \ref{subsec:time}. 

Temporal queries consist of keywords (in $Q$) and a set of temporal expressions. The temporal dimension in user input can be embedded in the textual part or not, which gives two types of queries \cite{gurrin_language_2010}:
\begin{itemize}
 \item  \textit{ Inclusive temporal queries}: the temporal dimension of the query is included in the keywords  (e.g: ``earthquake 17 august 1999''). In this case, the matching is done using standard IR techniques with the date tokens as keywords.
\item  \textit{ Exclusive temporal queries}:  the temporal dimension
of the query is distinct from its textual part  (e.g: ``earthquake''@ 08-17-1999) and is used to filter the results or in ranking function. In this paper, we use a strict filter -- i.e. a document either match or not the temporal filter given by the query, but there is no associated score.
\end{itemize}

\subsection{Temporal Aspects Model}
\label{sec:timewindows}

We now turn to the problem of estimating the set of temporal aspects given a term/query $t$, and the distribution $P(w|t)$, as needed by Equation \ref{eq:dcg_per_term}. This is the last part we need to define in order to perform static index pruning.

\begin{figure}[h!]
\centering
\includegraphics[scale=0.4]{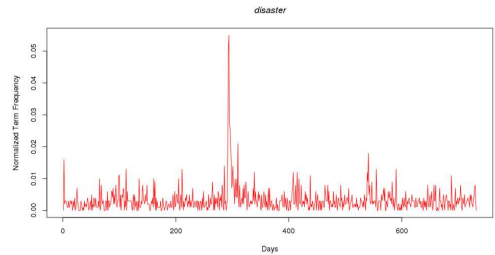}
\caption{Histogram for term 'disaster' in Los-Angeles Times (TREC)} \label{fig:exhistogram}
\end{figure}

For each term-query $t$ in the vocabulary of the document collection $D$, we can associate a time series, $S_t$  by identifying all the documents in the temporal collection that contain the term $t$. 
Then we take the sum of the term frequencies based on a chosen granularity (e.g: hour, day, week etc.) according to the documents temporal part $d_{time}$. 
In our work, we used the day granularity.
Figure \ref{fig:exhistogram} shows 
a time series generated for the term 'disaster' on Los Angles Times collection of TREC. 



\subsubsection {Fixed-sized Windows}

In this model, the time series of a term is partitioned into equal, fixed sized ($\gamma$) windows. 

Selecting the correct window size is an important issue for fixed-windows. When a small size is chosen, the number of total 
windows will increases and this becomes a bottleneck for big data collections. When a big size is chosen, with the decreasing number of windows, we will slowly start to diversify less. In setting the window size, we suppose that each term has a different time pattern. Hence, instead of using an uniform window size for all the terms, we propose to use an adaptive window size 
for each term, denoted $\gamma_t$. 

There have been many attempts to choose a good bin width for histograms like ours. 
Scott \cite{scott} proposed an approach based on the standard deviation, that was improved by 
Freedman-Diaconis \cite{freedman}. The latter work replaces the use of standard deviation with the \textit{interquartile range} (IQRN), the distance between the first and third quartile. The Freedman-Diaconis rule is known to be more robust to outliners, which is important for us to detect events in our timelines, and thus we adopted this method. Formally, we use the following window size for term $t$:

\[ \gamma_t = 2 IQRN^{-1/3} \]

\textbf{The simple window} strategy consists in using fixed sized non-overlapping windows. 
Formally,
 a window is represented as  $w_k = [s+k\times \gamma_t, s+(k+1)\times \gamma_t]$ where $s$ is an offset and $k\in\mathbb{Z}$.
We further assume that $P(w|t)$ follows uniform distribution, i.e. $ P(w | t) = \frac{1}{N} $ where $N$ is the number of non empty windows.

\vspace{.5em}
\textbf{The sliding window} strategy uses overlapping windows of fixed length. 
Formally, a window is represented as  $w_k = [s+k/2\times \gamma_t, s+(k/2+1)\times \gamma_t]$.
As in the previous model, we assume that $P(w|t)$ follows a uniform distribution.

\begin{figure}[h!]
\centering
\includegraphics[scale=0.4]{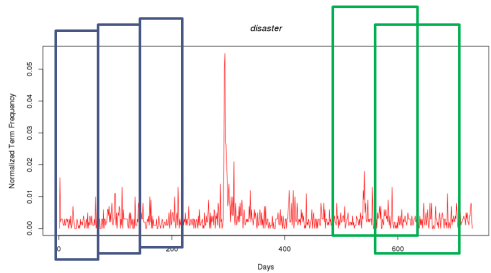}
\caption{Example for fixed windows} \label{fig:fixedwindows}
\end{figure}

\vspace{.5em}
Figure \ref{fig:fixedwindows} shows simple windows (in blue on the left) and sliding windows (in green on the right).

\subsubsection {Dynamic Windows}

In the above models, we considered windows of fixed size. However, events vary in duration. In the dynamic windows model we do not force the windows to have a fixed size but we detect the 
time windows based on the distribution of the documents over time. 

In order to do this, we use a probabilistic clustering based on Gaussian mixture models (GMM) of the data distribution.
This probabilistic model assumes that all the data points are generated from a mixture of a finite number of 
Gaussian distributions with unknown parameters. Each distribution can be thought of as a cluster.
The probability of generating a time stamp $x$ with the GMM is given \cite{bishop2006}:

\[p(x) = \sum_{k=1}^{K} \pi_k \mathcal{N} (x | \mu_k, \sigma_k^2) \] 
where the parameters $\pi_k$ are the mixing coefficients, which must sum to 1 and $\mathcal{N} (x | \mu_k, \sigma_k^2)$ is a Gaussian distribution defined by its mean $\mu_k$ and variance $\sigma_k^2 $.
%

A standard methodology consists in using the Expectation Maximization (EM) algorithm to estimate the finite the mixture models 
corresponding to each number of clusters considered and using the Bayesian Information Criteria (BIC) to select the number of mixture components, taken to be equal to the number of clusters \cite{bicfraley}.
The EM algorithm is an iterative refinement algorithm used for finding maximum likelihood estimates of parameters in probabilistic models. 

Choosing an appropriate number $K$ of clusters is essential for effective and efficient clustering. The standard way to estimate K is to start with one cluster and slowly increase the number
of clusters. At each K, the log-likelihood obtained from the GMM fit is used to compute BIC. The optimal value of K is the one with the smallest BIC. In this paper, this approach is used.

Finally, for each cluster with a mean $\mu$, variance $ \sigma^2$ and weight $\rho$, 
we associate a time window $ [\mu -\sigma,\mu +\sigma] $ with  $P(w|t) =  \rho$.

\subsubsection{Smoothing}

While we increase the importance of documents associated with several temporal aspects, 
we do not want to penalize too much highly relevant document associated to a few or only one aspects. 
In order to do this, we use a smoothing by defining a new window dubbed the \textit{Global window} $G$. 
This window contains spans all the time range of term $t$, i.e. all the documents belong to the global window. 
Following to this, we redefine the distribution of temporal aspects as follows:

\[ P(G|t) =  \lambda_w \]
\[ P^*(w|t) = (1 - \lambda_w) * P(w|t) \]

\begin{figure}[h!]
\centering
\includegraphics[scale=0.4]{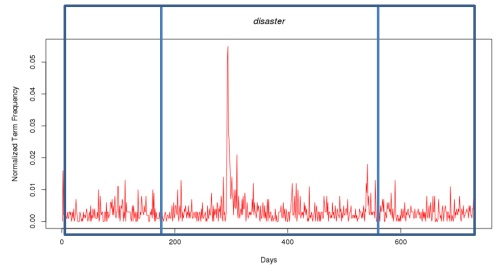}
\caption{Example for dynamic windows} \label{fig:dynamicwindows}
\end{figure}


\section{Experiments}
\label{sec:experiments}
In this section we present our experimental evaluation for the proposed approach. The experiments are conducted by using two publicly available collections.

Web archive collections are specific due to their temporal
dimension, so time must be present in the test collection (corpus, relevance
judgments etc.). However, there exists no standard dataset for our experiments because 
the existing ones do not contain time-related queries and the judgments are not targeted towards temporal information. 

Two publicly available document collections are used for our experiments. 
Both of them have time related information, but they differ in their properties, which make them complementary to evaluate our approach. 
The first one (LATimes) is a standard test collection for static index pruning experiments \cite{carmel_static_2001,CIKM_2012,blanco_static_2007,Blanco_2010}. This allowed us to ensure
that our results were in line with those reported in \cite{CIKM_2012,carmel_static_2001}.
LATimes contains time-stamped documents but no query with temporal dimension. 
The second collection (WIKI) is used in time-aware retrieval model tests. \cite{gurrin_language_2010, nattiyacompare}. 
It contains queries with time constraints and temporal expressions extracted from document but no time-stamped documents:

\begin{itemize}
 \item TREC Los Angeles Times (LATimes) \cite{trec} that contains approximately 132000 articles published in Los Angeles Times newspaper in 1989 and in 1990. 
We propose two different simulations to add 
temporal dimension to the collection: \textit{All relevant Time constraint} and \textit{Time constraint Test}
\item The English Wikipedia (WIKI) as of July 7,2009 that contains total of 2955294 encyclopedia articles. In this collection, the temporal expressions are extracted from the text and 
used as the validity interval of documents. 
\end{itemize}

In the next sections, we briefly present the collections, queries and relevance assessments used to evaluate the methods.

\vspace{.5em}\textbf{Baselines}

We compare our approaches \textit{Simple, Sliding, Dynamic} with the following baselines: the first one is the seminal work of Carmel \cite{carmel_static_2001} and the remaining are recent state-of-the-art work on index pruning.

\begin{itemize}
\item \textit{TCP - Static index pruning method presented in \cite{carmel_static_2001}}
The score of a posting is given by its TF-IDF value in the document, i.e. $tf \times \log(df/N)$ where $tf$ is the number of occurrences of the term in the document, $df$ is the number of documents where term occur, and $N$ is the number of documents in the collection. Then, for each term, the method selects the $k^{th}$ highest score, $z_t$, and sets the 
threshold value $\tau_t = \epsilon * z_t$, where $\epsilon$ is the parameter used to control the pruning rate. 
Each entry in the term posting list whose score is lower than predefined threshold $\tau_t$ is considered not important and 
removed from the posting list. 
Following \cite{CIKM_2012}, for TCP we set k = 10 to maximize the precision for the top 10 documents. 

\item \textit{IP-u - Static index pruning method presented in \cite{CIKM_2012}}
This approach is based on the notion of conditional entropy  of a document conditioned on a term:
\[ H (D|T) =\frac{1}{|T|}\sum_{t\in T}\sum_{d\in D} A (d,t)\]
where D is the set of documents and T is a set of terms, and
\[ A (d,t) =  -\frac{p (d|t)}{\sum_{d'}p (d'|t) } log\frac{p (d|t)}{\sum_{d'}p (d'|t) }\]

In their approach, they discard the postings $p_{t,d}$ whose contribution to the conditional
entropy $A(t,d)$ is lower than a given threshold.
We follow the original settings in \cite{CIKM_2012}, and use a Language Model with Jelineck-Mercer smoothing ($\lambda=0.6$) and use uniform document prior to estimate $p(d|t)$.

\item \textit{2N2P - Static index pruning method presented in \cite{Ecir_2011}}
2N2P uses a two-proportion Z-test, that compares differences between two random samples. 
In the context of pruning, the authors compare the distributions of a term
in the document and in the collection: if they differ too much, then it is necessary
to preserve the corresponding posting in the index, otherwise the collection approximation
is enough.

Formally, they compute the statistic
\begin{align*}
Z = \frac{\frac{tf_{t,d}}{|d|}-{\frac{tf_{t}}{|C|}}}{E} \mbox{ with } &
E = \sqrt{P (1-P)  ( \frac{1}{|d|} +  \frac{1}{|C|} )}
\\
&P = \frac{tf_{t,D} + ctf_{t} }{|d| +|C|}
\end{align*}
where $tf_{t,d}$ is the term frequency in the document $d$ of length $|d|$ and $ctf_{t}$ is the term frequency in
 the collection C of length (total number of term occurrences) $|C|$. $E$ corresponds to the standard deviation
All the postings with a
$Z$ score strictly lower than threshold $\epsilon$ are discarded.

\end{itemize}

\noindent
\textbf{Settings}

\noindent
Documents are indexed, retrieved and pruned
 using Apache Lucene\footnote{http://lucene.apache.org/}. 
 Our implementation does not update the document lengths after pruning.
The effectiveness is measured with mean average 
precision  (MAP) and normalized discounted cumulative gain (NDCG), two 
standard metrics in ad-hoc IR \cite{manning2008}. Both are biased towards top-ranked results,
but they exhibit a different behavior.

BM25 is used as the scoring function as follows.

 \[ score(q,d) = \sum_{i=1}^{|q|} idf(q_i) \frac{tf(q_i,d) (k_1 + 1)}{tf(q_i,d) + k_1 (1-b+b\frac{|d|}{avgdl})} \]

where 
\begin{itemize}
 \item $tf(q_i,d)$ correlates to the term's frequency, defined as the number of times that the query term qi appears in the document d.
\item $|d|$ is the length of the document d in words (terms). 
\item avgdl is the average document length over all the documents of the collection.
\item k1 and b are parameters, usually chosen as k1 = 2.0 and b = 0.75.
\item $idf(q_i)$ is the inverse document frequency weight of the query term $q_i$. It is computed by:

\[ idf(q_i) = log \frac{N-df(q_i) + 0.5}{df(q_i)+0.5}\]
where N is the total number of documents in the collection, and $df(q_i)$ is the number of documents containing the query term $q_i$. 
\end{itemize}

The pruning ratio reported in our results is the percentage of 
postings in the index removed by the algorithms. 
When the query contains a time constraint, we use it to filter out documents for which no time window intersect one of the query, i.e. $q_{time} \cap d_{time} = \emptyset $.\\

\begin{figure*}[t!]
      \begin{subfigure}{0.5\textwidth}
		\centering
                \includegraphics[scale=0.22]{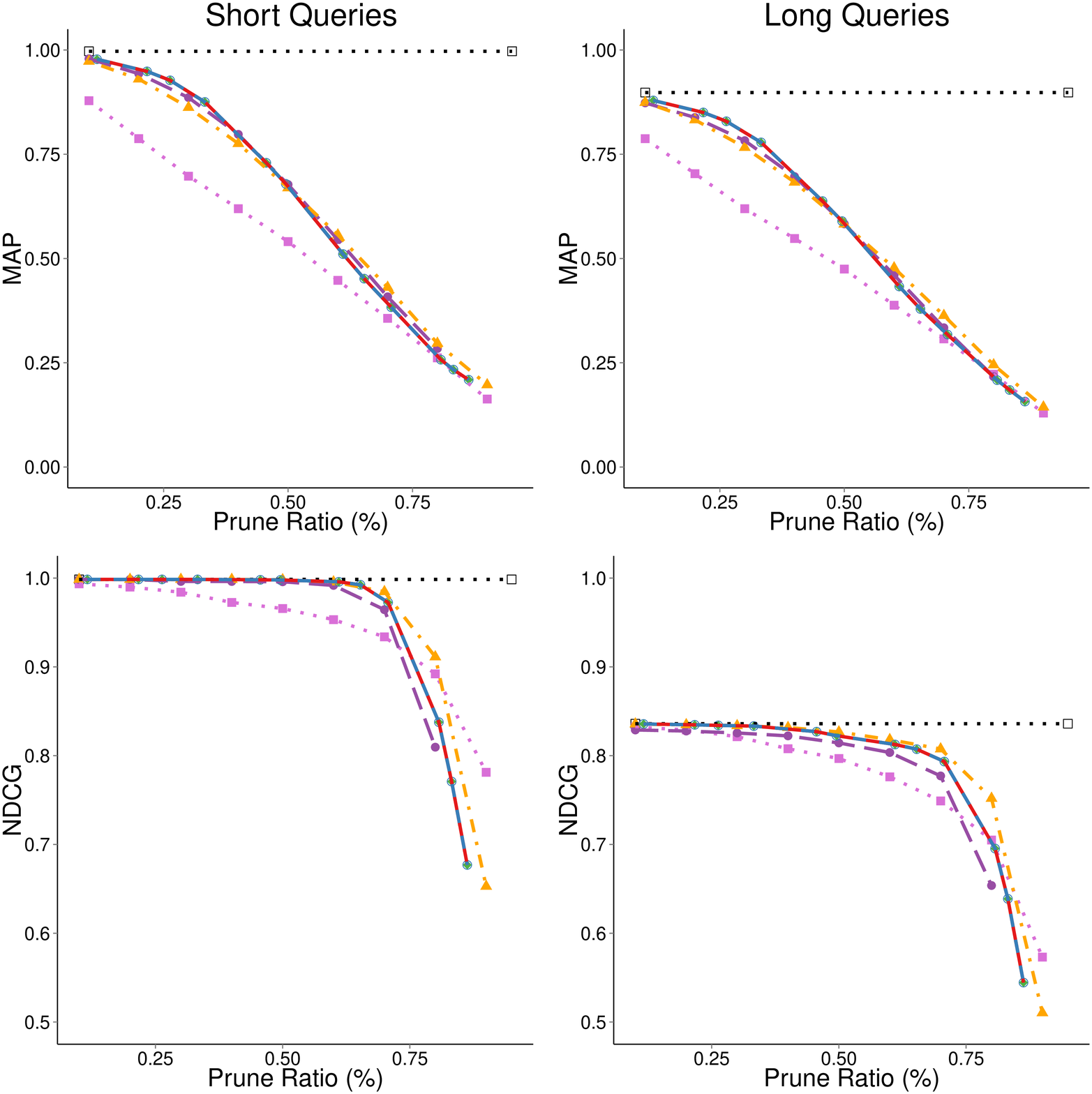}
		\caption{All relevant Time constraint Test results}
		\label{fig:trueLA}
        \end{subfigure}
      \begin{subfigure}{0.5\textwidth}
                \centering                
		\includegraphics[scale=0.22]{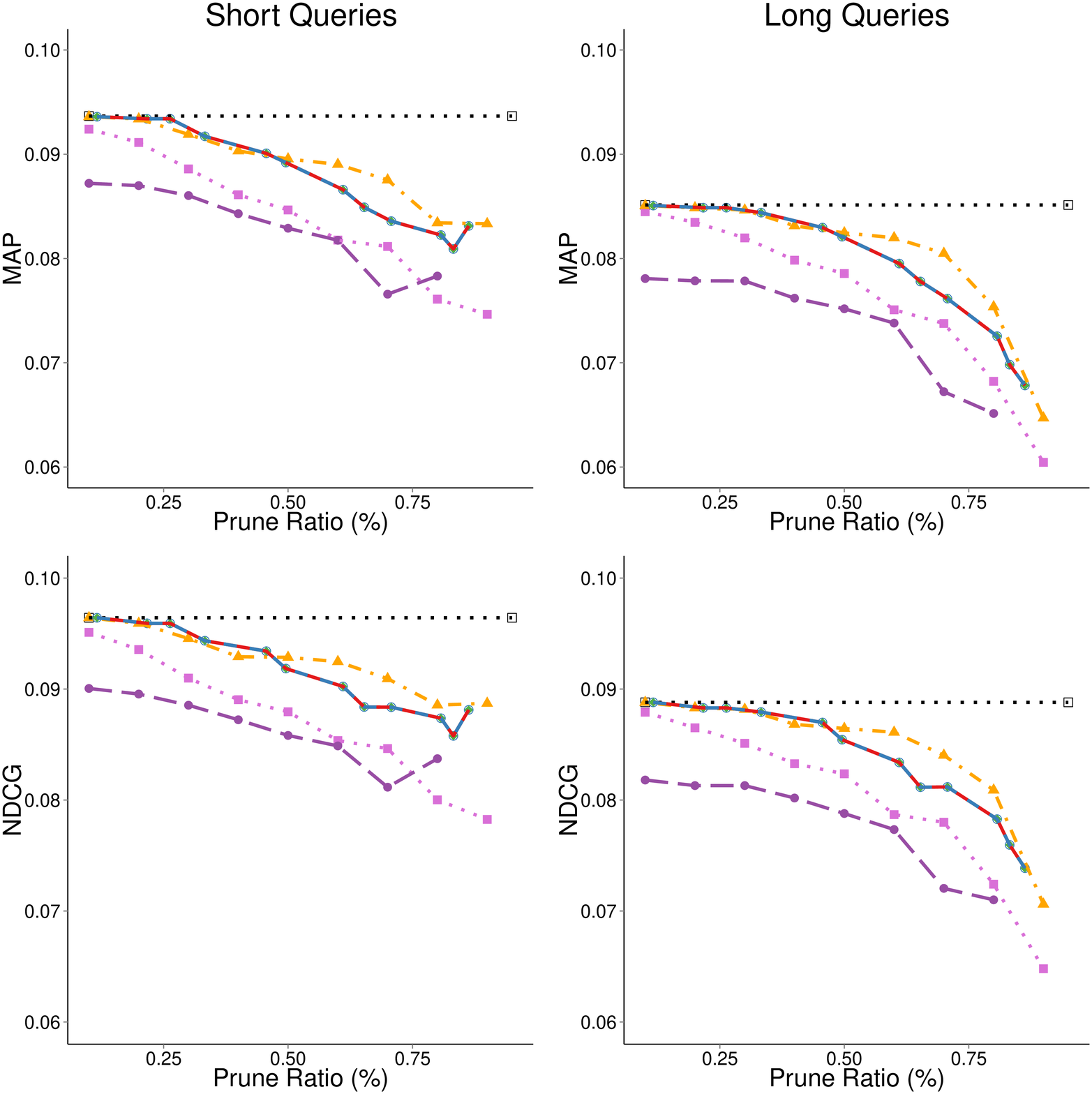}
 		\caption{Time constraint Test results} 		
		\label{fig:falseLA}
         \end{subfigure}
\\
      \begin{subfigure}{\textwidth}
                \centering                
		\includegraphics[scale=0.25]{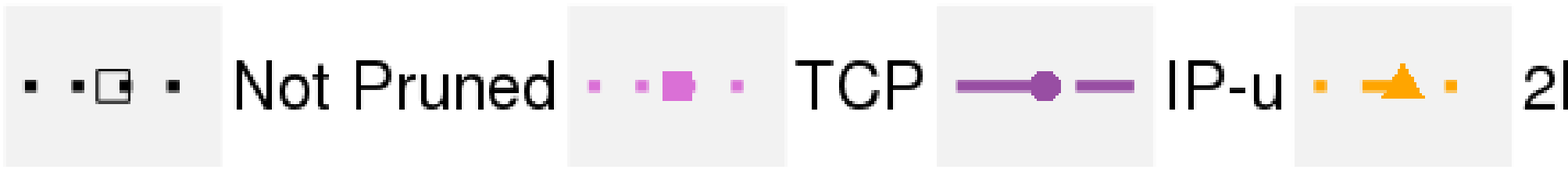}
         \end{subfigure}
    \caption{LATimes Temporal results}
\label{fig:latimes}
\end{figure*}

\subsection{ TREC Los-Angeles Times (LATimes) }

The TREC-LATimes dataset is a standard collection used for index pruning experiments. It consists of randomly 
selected articles published on the LA Times newspaper in 1989 and in 1990. 
It contains approximately 132000 articles. For the evaluation 
of this dataset we use TREC 6, 7 and 8 \textit{ad hoc} topics (topics 301-450).
Documents' publication dates were used as $d_{time}$, but as there
was no time associated with queries and their relevance judgements, 
we had to generate them from the original collection.

By using topics 301-450 and corresponding relevance judgments, we created two different 
datasets. For each topic, we randomly picked a time interval covered by the dataset 
and ran these temporal queries over initial index. The temporal 
queries that return at least one result with the non-pruned index were kept for our experiments.
By following this method, we obtained 1000 exclusive 
temporal short (title) and long (title+description) queries for different interval types: daily queries  $|q_{time}|$ = 1 day,
weekly queries $|q_{time}|$ = 7 days and monthly queries  $|q_{time}|$ = 30 days.
Experiments were conducted with daily, weekly and monthly queries, but this didn't lead
to changes in results -- we thus report weekly results only. 

The first test (\textit{All relevant Time constraint}) is conducted to check the behavior of our pruning methods. 
We consider any document from the specified time period containing at least one of the keywords to be relevant.

Figure \ref{fig:trueLA} plots MAP and NDCG measures as a function of the amount of pruning for weekly temporal queries. The first observation is that for all kind of queries TCP significantly performs worse than other methods. For
 queries with pruning 
ratio smaller than 50\%, our three proposed approaches performs slightly better than others. However, 2N2P does better at higher pruning level. 

\begin{figure}
\centering
\includegraphics[scale=0.22]{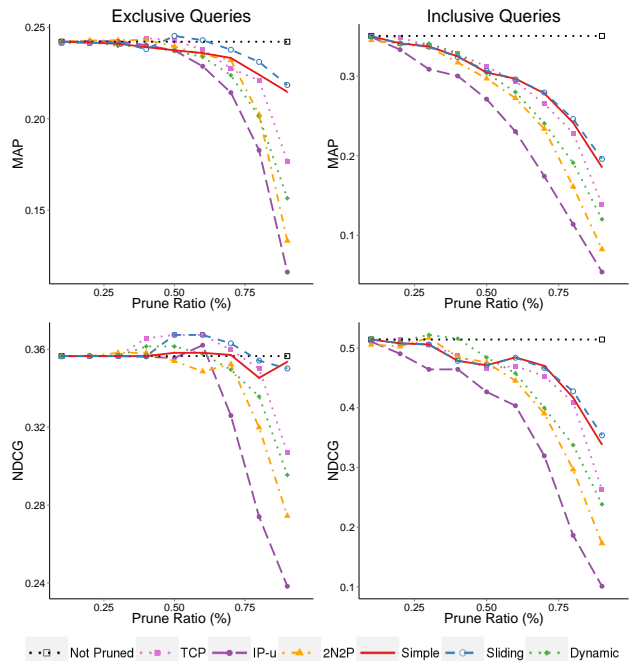}
\includegraphics[scale=0.20]{legend.eps}
\caption{WIKI overall results} \label{fig:all}
\end{figure}

In our second test, called \textit{Time constraint Test}, documents out of the time interval are considered as non-relevant but we kept the original relevance for the documents in the time interval. 
As shown in Figure \ref{fig:falseLA}, 
2N2P performs better at high pruning level for both long and short queries while our approaches is slightly better at low level pruning. The difference with respect to
 \textit{All related Time constraint Test}  is that 
IP-u stays behind TCP and the gap between 2N2P and our methods gets bigger at low pruning level.




As explained before, the temporal queries used in this section 
do not reflect the real temporal information needs of the users. 
More precisely, given the random generation of time intervals for queries, they do not
favor particularly our approach compared to the others. This is because
when the number of associated time intervals become high for each term, 
diversifying is not optimal.
The LATimes collection was used in preliminary experiments to confirm 
that our approaches perform similarly to the existing ones; the Wikipedia
corpus was then used to perform more realistic experiments, as described
in the next section.
 

\subsection{English Wikipedia Dump (WIKI) }                                                                                                                                                                             

Although there is an increasing interest in temporal collections and in research related to temporal information, to the best of our knowledge, there is only one available collection 
with temporal queries 
and corresponding relevance judgments which takes the temporal expressions in the documents into account. In \cite{gurrin_language_2010}, Berberich et al. used \textit{The English Wikipedia} 
dump of July 7, 2009 and obtained temporal queries for this collection 
by using Amazon Mechanical Turk (AMT). They selected 40 of those queries and categorized them 
according to their topic and temporal granularity (daily, weekly, etc.). 
The relevance judgments for these queries were also collected by using AMT 
and make them publicly available \footnote{http://www.mpi-inf.mpg.de/~kberberi/ecir2010/}.

Time constraints in queries refer explicitly to the time intervals that
the relevant wikipedia article refer to. 
In our experiments, each temporal expressions extracted from documents is considered
 as the validity interval of the documents and used to filter the results of temporal queries. 
We used 2330277 encyclopedia articles that contain temporal expressions, and, following  \cite{gurrin_language_2010},
considered  \textit{inclusive} and \textit{exclusive} queries as explained in Section \ref{sec:docqmodel}.

\begin{figure*}[ht!]
  \begin{subfigure}{0.5\textwidth}
		\centering
		\includegraphics[scale=0.22]{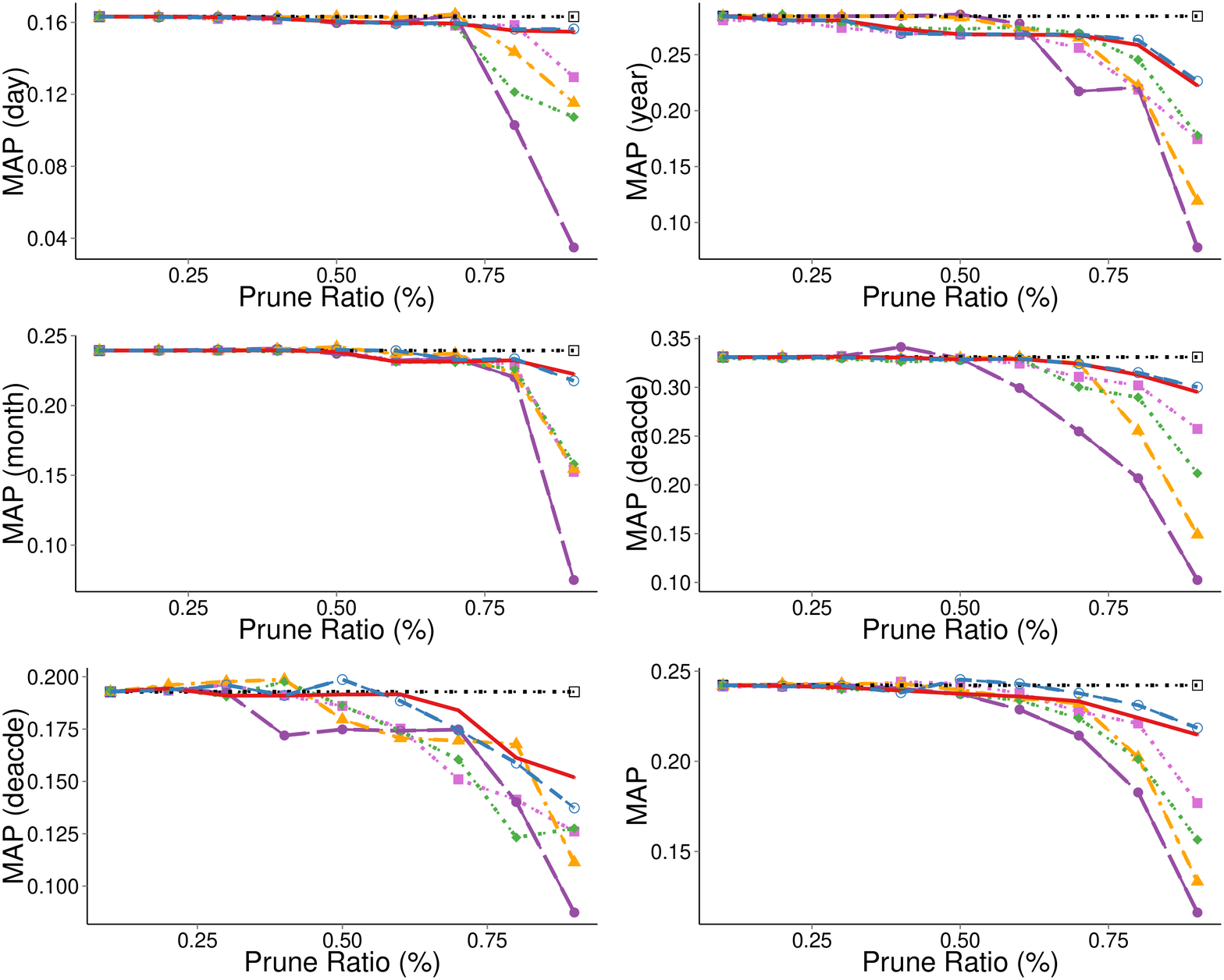}
		\caption{For Exclusive queries}
		\label{fig:mapbyqueryexcl}
        \end{subfigure}
      \begin{subfigure}{0.5\textwidth}
                  \centering
      \includegraphics[scale=0.22]{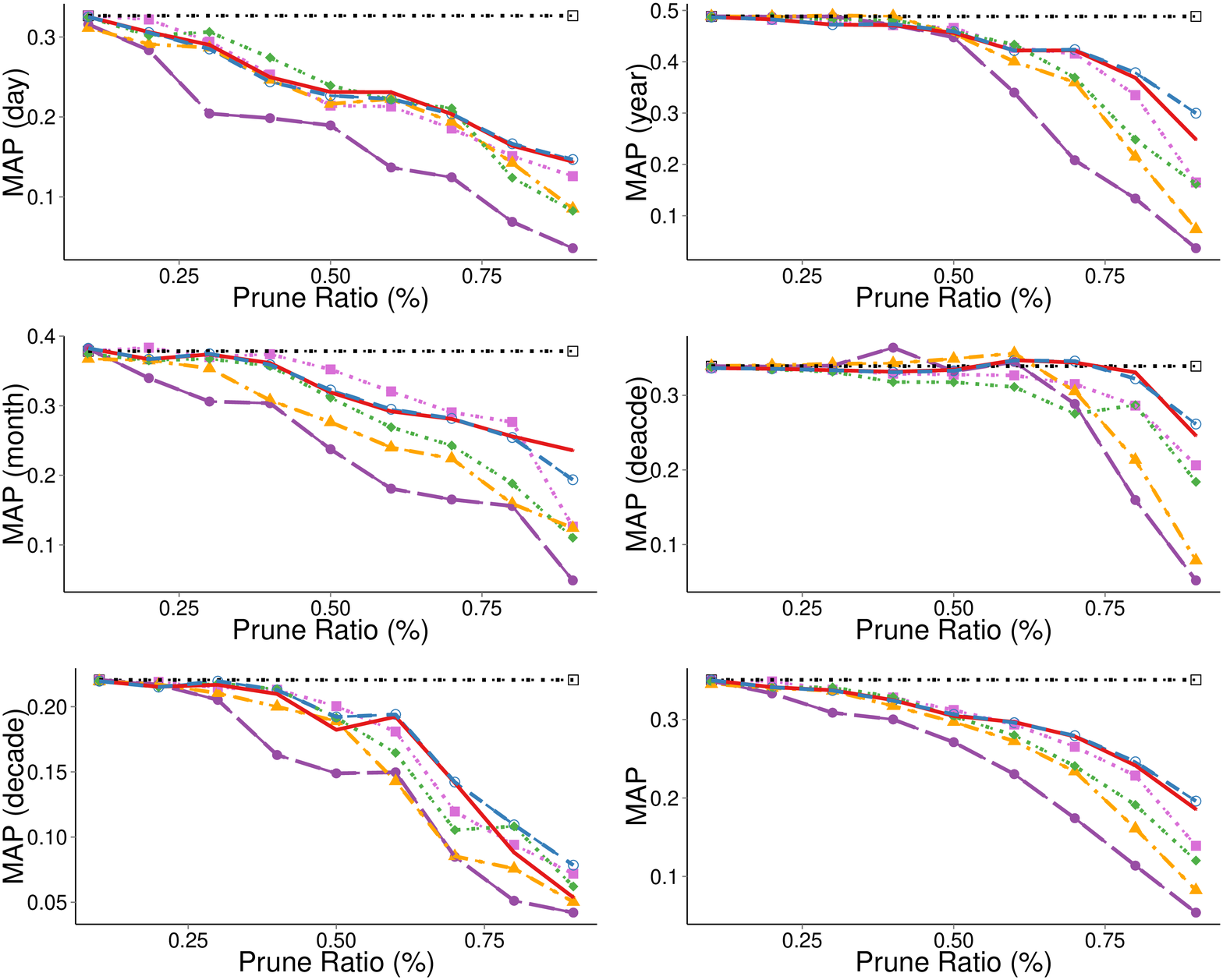}
      \caption{For Inclusive queries} \label{fig:mapbyqueryinc}
        \end{subfigure}
   \\
 \begin{subfigure}{\textwidth}
                \centering                
		\includegraphics[scale=0.30]{legend.eps}
         \end{subfigure}
    \caption{WIKI MAP by query time}
\label{fig:mapbyquery}
\end{figure*}

Figure \ref{fig:mapbyqueryexcl} plots MAP and NDCG as a function of the amount of pruning for inclusive and exclusive temporal queries. 
The results for initial index and the results obtained with topics 301-450 without temporal dimension are coherent with the baseline methods \cite{CIKM_2012,carmel_static_2001}. 
We can observe than that at low pruning levels, the performance can be higher for pruned indexes than with non pruned ones.
 This just shows that pruning removes postings that would have made non relevant documents retrievable.

We now turn to comparing the different pruning methods.
TCP, which performs worst with LATimes collections, significantly outperforms both IP-u and 2N2P.
Two of our methods, namely \emph{Simple} and \emph{Sliding}, perform
better than all the others whatever the query type \emph{inclusive} or \emph{exclusive},
and the gap between methods increases with the pruning level. This shows
that our time-based diversification method works well. 

However, our third method, based on dynamic windows, did not perform as well
contrarily to our expectations.
It can be related to number of clusters chosen as input for GMM as discussed in Section \ref{sec:tsip}. The clusters obtained did not cover all the documents for some terms - in this case, we chose
the closest cluster.
In our future works, we plan
to investigate further how dynamic windows can be defined, in particular 
by using other clustering algorithms.

We also investigated the behavior of the pruning methods for queries
referring to different temporal intervals (Day, Month, Year, Decade, Century). The
dataset was constructed so that each group contains 8 queries.
Figure \ref{fig:mapbyquery} shows the MAP values for exclusive and inclusive queries.
We can see that whatever the temporal interval, our proposed approach perform better, but the
gap decreases with higher temporal intervals (like year or decade). This is coherent with our approach
since when the temporal interval covers a larger time span, diversifying the time aspect
of the documents become less important.




\section{Conclusion and Discussion}
\label{sec:concl}

We studied the problem of search result diversification in the context of index pruning. We proposed a new approach called diversification based static index pruning that decides which postings to prune by
taking into account both the (estimated) relevance of the documents and the diversification of the results: For each term, we preserve the top-$k$ postings that maximize a chosen IR metric (DCG in this paper) using a greedy algorithm, and discard the rest. In order to perform this pruning efficiently, we proposed an algorithm for selecting the next best posting to preserve.

Our original motivation was to explore index pruning strategies for temporal collections, since these collections are characterized by large index sizes. We thus applied our diversification-based index pruning method to
take into account the temporal dimension. This was achieved by associating to each possible term-query a distribution over time windows (time range with a start and end date), based on the time windows associated to each of the documents
containing the term at hand. 
We proposed three different approaches to compute namely Simple, Sliding and Dynamic windows. 

Our experiments were conducted on two publicly available collections: LATimes and WIKI. 
The LATimes collection does not contain the temporal information needs, and was used
to show that our approach is competitive with state-of-the-art pruning algorithms.
The experiments on WIKI showed that Simple and Sliding methods perform better overall for temporal
queries, but we were unable to show good results with the more sophisticated dynamic
strategy.

This work is the first one dealing with the (temporal) diversification based index pruning.
As such, many different extensions and enhancements of our approach are possible and will be
explored in future work.

First, we can explore a better way to estimate the distribution over queries by using query logs, or term co-occurrence information - this would exploit the redundant information that might be contained in two or more terms.

Second, we can explore the behavior of index pruning with more sophisticated time-aware
ranking models like those exposed  in \cite{gurrin_language_2010}, where the matching between
time intervals is not boolean anymore but is real-valued. This approximate matching could be
exploited to reduce further the index size.

Finally, one of the challenges in the context of web archives, besides the temporal dimension, is redundancy: 
Different versions of the same document can be present in the document collections. The changes between the versions are likely to be small for most of documents, 
and this should be exploited by keeping only a few (ideally one) of the postings (for the same term) corresponding to the different versions of the document. 
This would however require the development of a specific test collection based on web archive crawls with related temporal queries and relevance judgements.

\bibliographystyle{abbrv} 
\bibliography{Pruning2} 

\end{document}